\documentclass[onecollarge,natbib]{svjour2}
\bibpunct{[}{]}{;}{n}{}{,} % to get "[numbered]" references from natbib

\usepackage{graphicx}
\usepackage{mathptmx}

\usepackage{amsmath}
\usepackage{feynmp}
\usepackage{tikz}
\usepackage{units}

\journalname{Few-Body Systems (EFB23)}

\title{Electric Properties of One-Neutron Halo Nuclei in Halo EFT}
\subtitle{}

\author{Jonas Braun \and Hans-Werner Hammer}

\institute{J. Braun 
    \at  {Institut f\"ur Kernphysik,
Technische Universit\"at Darmstadt, 64289 Darmstadt, Germany}\\
      \email{braun@theorie.ikp.physik.tu-darmstadt.de} 
    \and H.-W. Hammer 
\at{Institut f\"ur Kernphysik,
Technische Universit\"at Darmstadt, 64289 Darmstadt, Germany}
\at{ExtreMe Matter Institute EMMI, GSI Helmholtzzentrum
f\"ur Schwerionenforschung GmbH, 64291 Darmstadt, Germany}
}

\date{Received: date / Accepted: date}

\begin{document}
 \maketitle
 
 \begin{abstract} We exploit the separation of scales in weakly-bound nuclei to compute E2 transitions and electric form factors in a Halo EFT framework. The relevant degrees of freedom are the core and the halo neutron. The EFT expansion is carried out in powers of $R_{core}/R_{halo}$, where $R_{core}$ and $R_{halo}$ denote the length scales of the core and halo, respectively.
 We include the strong $s$-wave and $d$-wave interactions by introducing dimer fields. The dimer propagators are regulated by employing the power divergence subtraction scheme and matched to the effective range expansion in the respective channel.
 Electromagnetic interactions are included via minimal substitution in the Lagrangian. We demonstrate that, depending on the observable and respective partial wave, additional local gauge-invariant operators contribute in LO, NLO and higher orders.
 
 $*$ This work has been supported by Deutsche Forschungsgemeinschaft (SFB 1245).

 \end{abstract}

  \section{Introduction}
  \label{intro}
  Electric properties provide a unique window on the structure and dynamics of one-neutron halo nuclei. Two prominent examples of one-neutron halo nuclei are $^{11}$Be and $^{15}$C~\cite{jensen2004structure}. Due to their weakly-bound nature, the length scales $R_{halo}$ of the halo and $R_{core}$ of the core are separated. As a consequence, observables can be calculated using Halo EFT~\cite{bertulani2002effective,bedaque2003narrow} which exploits this separation of scales between $R_{core}$ and $R_{halo}$. The Halo EFT formalism has been successfully used to study various reactions and properties of halo-like systems.
  Some examples include $s$-wave $\alpha\alpha$ resonance in ${}^8$Be~\cite{higa2008alphaalpha}, universal properties of two-neutron halo nuclei~\cite{canham2008universal, canham2010range}, radiative neutron/proton capture on ${}^7$Li~\cite{rupak2011model,zhang2014combiningLi} and on ${}^7$Be~\cite{zhang2014combiningBe} and the ground state of ${}^{19}$C~\cite{acharya201319}. In this work, we follow the approach presented in Ref.~\cite{hammer2011electric}, where electric properties of ${}^{11}$Be are calculated using Halo EFT. The $\frac{1}{2}^+$ ground state of ${}^{11}$Be is predominantly an $s$-wave bound state and the $\frac{1}{2}^-$ first excited state predominantly a $p$-wave bound state.

  We apply the Halo EFT framework to ${}^{15}$C, which has also a $\frac{1}{2}^+$ ground state. In contrast to ${}^{11}$Be, the first excited $\frac{5}{2}^+$ state of ${}^{15}$C is predominantly a $d$-wave bound state. We include the strong $d$-wave interaction by introducing a new dimer field and compute electric form factors and the E2 transition strength. In a similar way, electromagnetic form factors for the ground state of ${}^{15}$C and the radiative neutron capture on ${}^{14}$C to the $\frac{1}{2}^+$ ground state of ${}^{15}$C are calculated in Refs.~\cite{fernando2015electromagnetic, rupak2012radiative}, respectively.

  \section{Halo EFT formalism}
  \label{formalism}
   The neutron separation energy of the $\frac{1}{2}^+$ state of ${}^{15}$C is 1218 keV and the neutron separation energy of the $\frac{5}{2}^+$ state is 478 keV~\cite{ajzenberg1991energy}. The first excitation of the ${}^{14}$C nucleus is 6.1 MeV above the $0^+$ ground state. Compared to the bound states of ${}^{14}$C the weakly-bound states of ${}^{15}$C indicate their one-neutron halo nature. 
   Converting these energy scales into the relevant distance scales $R_{core}$ and $R_{halo}$ in this system we can carry out our EFT expansion in powers of $R_{core}/R_{halo} \approx 0.3-0.45$. Since the ratio is not particularly small, effects beyond LO are not negligible in our theory.
   
   We follow the Halo EFT formalism developed in Ref.~\cite{hammer2011electric}. The relevant degrees of freedom are the ${}^{14}$C core, represented by a bosonic field $c$, and the halo neutron, represented by a spinor field $n$. The strong $s$-wave and $d$-wave interactions are included through the incorporation of auxiliary spinor fields $\sigma_s$ and $d_{m}$, which correspond to the $\frac{1}{2}^+$ and $\frac{5}{2}^+$ states, respectively. With the convention that repeated spin indices are summed, the non-relativistic Lagrangian is written as:
  \begin{align}
  \notag
    \mathcal{L} = \ & c^\dagger \left(i \partial_t + \frac{\nabla^2}{2M}\right) c + n^\dagger \left(i \partial_t + \frac{\nabla^2}{2m}\right) n \\ \notag
    & + \sigma_s^\dagger \left[\eta_0 \left(i \partial_t + \frac{\nabla^2}{2M_{nc}}\right) + \Delta_0 \right] \sigma_s + d_{m}^\dagger \left[c_2 \left(i \partial_t + \frac{\nabla^2}{2M_{nc}}\right)^2 + \eta_2 \left(i \partial_t + \frac{\nabla^2}{2M_{nc}}\right) + \Delta_2 \right] d_{m}\\
    \label{eq:lagrangian}
    & - g_0 \left[c^\dagger n_s^\dagger \sigma_s + \sigma_s^\dagger n_s c\right] - g_2 \left[d^\dagger_{m} \left[n \overset{\leftrightarrow}\nabla^2 c\right]_{\frac{5}{2},m} + \left[c^\dagger \overset{\leftrightarrow}\nabla^2 n^\dagger \right]_{\frac{5}{2},m} d_{m} \right] \ ,
  \end{align}
  where $\overset{\leftrightarrow}\nabla$ is the Galilean invariant derivative, $m$ denotes the neutron mass, $M$ the core mass and $M_{nc} = m + M$ is the total mass of the n-${}^{14}$C system and we project out the $J=5/2$ part of the strong $d$-wave interaction.

  \subsection{Dressing the $d$-wave state}
  \begin{figure}[t]
    \centering
    \includegraphics[width=0.99\textwidth]{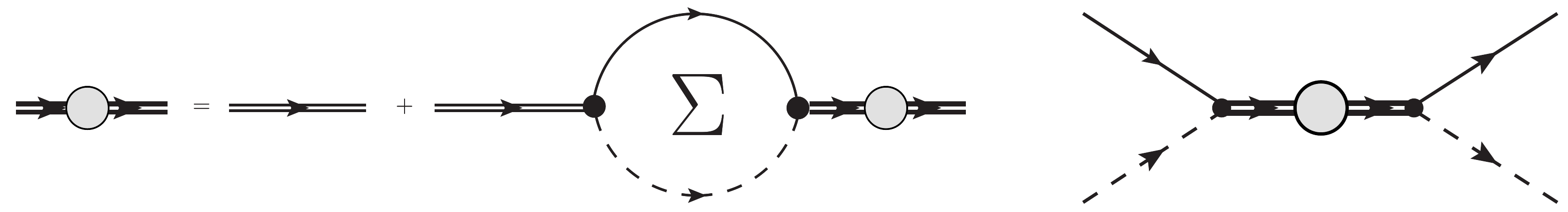}
    \caption{The dashed line denotes the core field $c$, and the thin, solid line the neutron. The thin double line represents the bare dimer propagator, and the thick double line with the gray circle is the dressed dimer propagator. The left panel shows the diagrammatic representation of the Dyson equation for the dressed dimer propagator and the right panel the neutron-core scattering amplitude with the dressed dimer propagator.}
    \label{fig:dimer-propagator}
  \end{figure}
  The dressed $d$ propagator and the $d$-wave scattering amplitude are computed from the diagrams in Fig. \ref{fig:dimer-propagator} in the same way as in Ref.~\cite{hammer2011electric}.
  The Dyson equation for the $d$-wave is illustrated in the left panel of Fig. \ref{fig:dimer-propagator} with the one-loop self-energy $\Sigma_d(p)$. We use the power divergence subtraction (PDS) as regularization scheme~\cite{kaplan1998new, kaplan1998two}.
  The $d$-wave scattering amplitude in the two-body center-of-mass frame with $E=k^2/(2m_R)$ and $k=|\mathbf{p}'|=|\mathbf{p}|$ for on-shell scattering is matched to the effective range expansion (ERE).
  Since we have to renormalize arising divergences, the effective range parameters $a_2, r_2$ and $\mathcal{P}_2$ are required at LO, which leads to the power-counting scheme below. Furthermore, we need the additional higher order kinetic term with coupling constant $c_2$ for the bare propagator in our Lagrangian \eqref{eq:lagrangian}~\cite{beane2001rearranging}.
  
  The assumed power counting for the $D_p(p)$ propagator follows the ideas of Refs.~\cite{bedaque2003narrow, hammer2011electric, rupak2012radiative}.
  We apply the constraint to have the minimal number of fine tunings possible in our power-counting scheme. This requirement is motivated by the fact that every additional fine tuning makes our scenario less likely to be found in nature, as discussed in Ref.~\cite{bedaque2003narrow}. However, the number of fine tunings for higher partial waves $(l>1)$ is larger than proposed in Ref.~\cite{bedaque2003narrow}. Since arising divergences in the calculation of the dimer propagator have to be renormalized properly, the number of ERE parameters needed at LO for the $l$-th partial wave is $(l+1)$.
  Thus, for the $l$-th partial wave and $l \geq 1$ we need $l$ fine-tuned parameters. This suggests that higher partial wave bound states are less likely to appear in nature which is consistent with experimental observations.
  In the case of the $D_d(p)$ propagator, two out of three ERE parameters need to be fine-tuned, because $a_2 \sim R_{halo}^4 \ R_{core}$ and $r_2 \sim 1/(R_{halo}^2 \ R_{core})$ are both unnaturally large. This scenario is more likely to occur than three unnaturally large parameters and three fine-tunings such that $\mathcal{P}_2 \sim 1/R_{core}$. Higher ERE terms are suppressed by powers of $R_{core}/R_{halo}$.
  Thus, the relevant fit parameters in our EFT at LO are $\gamma_0$, $\gamma_2$, $r_2$, and $\mathcal{P}_2$.

  \subsection{Electromagnetic interactions}
  Electromagnetic interactions are included via minimal substitution in the Lagrangian, $\partial_\mu \rightarrow D_\mu = \partial_\mu + i e\hat{Q} A_\mu$. Additionally, we have to consider further, local, gauge-invariant operators involving the electric field $\mathbf{E}$ and the fields $c$, $n$, $\sigma$ and $d$. Depending on the observable and respective partial wave they contribute at different orders in our EFT.

  \section{Results}
  
  \subsection{Electric form factors}
  \begin{figure}[t]
    \centering
    \includegraphics[width=0.95\textwidth]{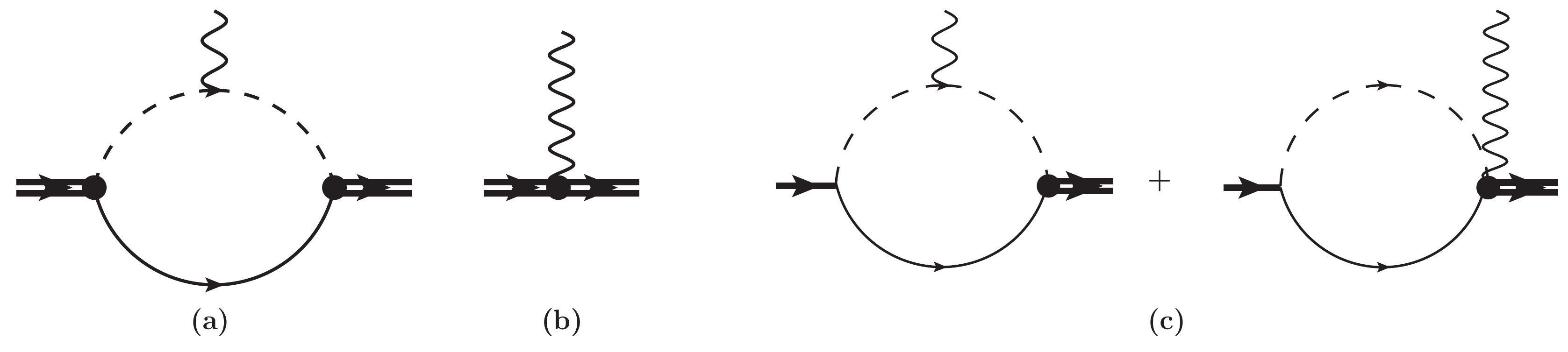}
    \caption{The diagrams (a) and (b) contributing to the irreducible vertex for an $A_0$ photon to couple to the ${}^{14}$C-neutron $d$-wave bound state field up to NLO. 
    Diagram (a) emerges from the minimal substitution in the strong $d$-wave interaction and contributes at NLO while diagram (b) arises since $r_2$, $\mathcal{P}_2$ and the local, gauge-invariant operator $\sim L_{C01/2}^{(d)}$ contribute at LO.
    Diagram (c): The two LO diagrams contributing to the irreducible vertex that determines the $s$-to-$d$ state transition in Halo EFT.}
    \label{fig:formfactorsE2}
  \end{figure}
  The result for the electric form factor of the $s$-wave bound state is given in Ref.~\cite{hammer2011electric} for ${}^{11}$Be and in Ref.~\cite{fernando2015electromagnetic} for ${}^{15}$C. We follow the approach of the first reference and compute the $d$-wave form factors by calculating the contribution to the irreducible vertex for $A_0 d d$ interactions up to NLO shown in Fig. \ref{fig:formfactorsE2}.
  The computation is carried out in the Breit frame, where $q = (0,\mathbf{q})$, and the irreducible vertex for the $A_0$ photon coupling to the $d$ state in Cartesian coordinates yields:
  \begin{align}
    \Big \langle ij \left| J^0 \right| op \Big \rangle =& - i e Q_c \left[ G_E(|\mathbf{q}|) \ E_{ij,op} + \frac{1}{2 M_{nc}^2} G_Q(|\mathbf{q}|) \ Q_{ij,op} + \frac{1}{4 M_{nc}^4} G_H(|\mathbf{q}|) \ H_{ij,op} \right] \ ,
  \end{align}
  with the three momentum of the virtual photon $\mathbf{q}=\mathbf{p'}-\mathbf{p}$ and three different $d$-wave tensors for each form factor $E_{ij,op} \sim q^0$, $Q_{ij,op} \sim q^2$ and $H_{ij,op} \sim q^4$.
  The neutron spin is unaffected by the charge operator up to the order considered here and we get results for the electric $G_E(|\mathbf{q}|)$, quadrupole $G_Q(|\mathbf{q}|)$ and hexadecupole $G_H(|\mathbf{q}|)$ form factors. The local, gauge-invariant operator $\sim L_{C01/2}^{(d)}$ is needed at LO to counter the arising divergences for the charge radius and quadrupole moment. At LO we obtain the following results for the charge and quadrupole radii and the quadrupole and hexadecupole moment:\\ \vspace{-1em}
  \begin{align}
    \langle r_E^2 \rangle^{(d)} \ =  \frac{-6 \tilde{L}_{C0E}^{(d) \text{ LO}}}{r_2 + \mathcal{P}_2 \gamma_2^2} , \ \
    \langle r_Q^2 \rangle^{(d)} = \frac{90 f^4}{7 \gamma_2 \left(r_2 + \mathcal{P}_2 \gamma_2^2\right)} , \ \
    \mu_Q^{(d)} =  \frac{40 \tilde{L}_{C02}^{(d)\text{ LO}}}{3 \left(r_2 + \mathcal{P}_2 \gamma_2^2 \right)} , \ \
    \mu_H^{(d)} = \frac{-2 f^4}{3 \gamma_2 \left(r_2 + \mathcal{P}_2 \gamma_2^2\right)} ,
  \end{align}
  with $f = m_R/M$.
  Since the quadrupole radius $\langle r_Q^2 \rangle^{(d)}$ and the hexadecupole moment $\mu_H^{(d)}$ differ only by a constant factor at LO, we obtain a smooth correlation between both observables $\langle r_Q^2 \rangle^{(d)} = -\frac{135}{7} \mu_H^{(d)}$.

  \subsection{E2 transition}

  The diagrams contributing to the irreducible vertex for the E2 transition from the $1/2^+$ state to the $5/2^+$ state at LO are shown in Fig. \ref{fig:formfactorsE2}.
  The photon has a four momentum of $k=(\omega, \mathbf{k})$ and its polarization is denoted by $\mu$.
  The computation of both diagrams yields a vertex function $\Gamma_{m' s \mu}$, where $m'$ is the total angular momentum projection of the $5/2^+$ state and $s$ denotes the spin projection of the $1/2^+$ state. 
  We calculate the irreducible vertex in Coulomb gauge such that we have $\mathbf{k} \cdot \epsilon = 0$ for real photons and obtain for B(E2):
  \begin{align}
  \notag
  \text{B(E2)} = \frac{6}{15 \pi} \ Z_{eff}^2 e^2 \ \frac{\gamma_0}{-r_2 -\mathcal{P}_2 \gamma_2^2} \ \left[ \frac{3\gamma_0^2 + 9\gamma_0\gamma_2 + 8\gamma_2^2}{(\gamma_0 + \gamma_2)^3} \right]^2 \ ,
  \end{align}
  where the effective charge is given by $Z_{eff} = \left(\frac{m}{M_{nc}}\right)^2 Q_c$~\cite{typel2005electromagnetic}.

  The experimental result for the E2 transition strength is $\text{B(E2)}_\text{W.u.} = 0.44 (1)$ and thus, $\text{B(E2)} = 0.97 (2) \ e^2 \text{fm}^{4}$.
  It follows that $1/\left(r_2 + \mathcal{P}_2 \gamma_2^2\right) = -323(7) \ \text{fm}^{3}$ and considering the previous section we can predict $\mu_H^{(d)} = 0.030 \ e\text{fm}^{4}$, $\langle r_Q^2 \rangle^{(d)} = -0.578 \ \text{fm}^{4}$, and $\langle r_H^2 \rangle^{(d)} = 0.004 \ \text{fm}^{6}$.
  
  In the future, we plan to investigate the correlations found in Halo EFT in \textit{ab initio} calculations.

% \bibliography{dwave-E2}
% \bibliographystyle{spbasic}

\end{document}